\documentclass[12pt,preprint]{aastex}

\received{}
\accepted{}
\journalid{VOL}{JOURNAL DATE}
\articleid{START PAGE}{END PAGE}
\paperid{MANUSCRIPT ID}

\lefthead{Yoshida \& Lee}
\righthead{Relativistic $r$-modes in Slowly Rotating Neutron Stars}

\begin{document}

%%%%%%%%%%%%%%%%%%%%%%%%%%%%%%%%%%%%%%%%%%%%%%%%%%%%%%%%%%%%%%
%% O. FRONT MATTER                                          %%
%%%%%%%%%%%%%%%%%%%%%%%%%%%%%%%%%%%%%%%%%%%%%%%%%%%%%%%%%%%%%%

\title{Relativistic $r$-modes in Slowly Rotating Neutron Stars: \\
           Numerical Analysis in the Cowling Approximation}
\author{Shijun Yoshida\altaffilmark{1} and Umin Lee}
\affil{Astronomical Institute, Graduate School of Science, 
Tohoku University, Sendai 980-8578, 
Japan \\ yoshida@astr.tohoku.ac.jp, lee@astr.tohoku.ac.jp}

\altaffiltext{1}{Research Fellow of the Japan Society for 
the Promotion of Science.}

\begin{abstract}

We investigate the properties of relativistic $r$-modes of slowly rotating 
neutron stars by using a relativistic version of the Cowling 
approximation. 
In our formalism, we take into account the influence of the Coriolis like 
force on the stellar oscillations, but ignore the effects of 
the centrifugal like force. For three neutron star models,
we calculated the fundamental $r$-modes with $l'=m=2$ and $3$.
We found that the oscillation frequency $\bar\sigma$ of
the fundamental $r$-mode is in a good approximation given by 
$\bar\sigma\approx \kappa_0\,\Omega$, where 
$\bar\sigma$ is defined in the 
corotating frame at the spatial infinity, and
$\Omega$ is the angular frequency of rotation of the star.
The proportional coefficient $\kappa_0$ is only weakly dependent on $\Omega$, but
it strongly depends on the relativistic parameter $GM/c^2R$, where $M$ and $R$ are 
the mass and the radius of the star.
All the fundamental $r$-modes with $l'=m$ computed in this study
are discrete modes with distinct regular eigenfunctions, and
they all fall 
in the continuous part of the frequency spectrum associated with
Kojima's equation (Kojima 1998).
These relativistic $r$-modes are
obtained by including the effects of rotation higher
than the first order of $\Omega$ so that the buoyant force plays a role, 
the situation of which is quite similar to that
for the Newtonian $r$-modes.

\end{abstract}


%%%%%%%%%%%%%%%%%%%%%%%%%%%%%%%%%%%%%%%%%%%%%%%%%%%%%%%%%%%%%%
%%  INTRODUCTION                                            %%
%%%%%%%%%%%%%%%%%%%%%%%%%%%%%%%%%%%%%%%%%%%%%%%%%%%%%%%%%%%%%%

\section{Introduction}

It is Andersson (1998) and Friedman \& Morsink (1998) who realized 
that the $r$-modes in rotating stars are unstable against 
the gravitational radiation reaction.
Since then, a large number of papers have been published to explore
the possible importance of the instability in neutron stars.
The $r$-modes in rotating stars are restored by the Coriolis force.
They have the dominant toroidal component 
of the displacement vector and their oscillation frequencies are
comparable to the rotation frequency $\Omega$ of the star
(see, e.g., Bryan 1889; Papaloizou \& Pringle 1978;
Provost, Berthomieu, \& Roca 1981; Saio 1982; Unno et al. 1989).
For the $r$-mode instability,
see reviews by, e.g., Friedman \& Lockitch (1999),
Andersson \& Kokkotas (2001), Lindblom (2001), and
Friedman \& Lockitch (2001).
Most of the studies on the $r$-mode instability in neutron stars, however,
have been done within the framework of Newtonian dynamics.
Since the relativistic factor can be as large as $GM/c^2R\sim 0.2$
for neutron stars where $M$ and $R$ are respectively the mass and the radius, 
the relativistic effects on the $r$-modes are essential.

From time to time the effects of general relativity on the $r$-modes
in neutron stars have been discussed in the context of the $r$-mode instability.
Kojima (1998) is the first who investigated the $r$-modes in neutron stars
within the framework of general relativity.
In the slow rotation approximation, 
Kojima (1998) derived a second order ordinary 
differential equation governing the relativistic $r$-modes, expanding the 
linearized Einstein equation to the first order of $\Omega$, and assuming that 
the toroidal component of the displacement vector is dominant and 
the oscillation frequency is comparable to $\Omega$.
Kojima (1998) showed that this equation has a singular 
property and allows a continuous part in the frequency spectrum
of the $r$-modes
(see, also, Beyer \& Kokkotas 1998). 
Recently, Lockitch, Andersson, \& Friedman (2001) showed that Kojima's equation 
is appropriate for non-barotropic stars but not for barotropic ones. 
They found that both discrete regular $r$-modes and continuous singular 
$r$-modes are allowed in Kojima's equation for uniform density stars, and 
suggested that the discrete regular $r$-modes are a relativistic 
counterpart of the Newtonian $r$-modes. 
Yoshida (2001) and Ruoff \& Kokkotas (2001a) 
showed that discrete regular $r$-mode solutions to Kojima's equation
exist only for some restricted ranges of the polytropic index and the relativistic
factor for polytropic models, and 
that regular $r$-mode solutions do not exist for the typical ranges of the
parameters appropriate for neutron stars.

The appearance of continuous singular $r$-mode solutions in
Kojima's equation (Kojima 1998) caused a stir in the community of 
people who are interested in the $r$-mode instability.
Lockitch et al. (2001) (see also Beyer \& Kokkotas 1999) suggested that 
the singular property in Kojima's equation could be avoided if 
the energy dissipation associated with the gravitational radiation is
properly included in the
eigenvalue problem because the eigenfrequencies become complex due to the
dissipations
and the singular point can be detoured when integrated along the real axis.
Very recently, this possibility of avoiding the singular property in 
Kojima's equation has been examined by Yoshida \& Futamase 
(2001) and Ruoff \& Kokkotas (2001b), who showed that the basic 
properties of Kojima's equation do not change even if the gravitational 
radiation reaction effects are approximately included into the equation. 
Quite interestingly, 
as shown by Kojima \& Hosonuma (2000), if the third order rotational effects are added
to the original Kojima's equation (Kojima 1998),
the equation for the $r$-modes becomes 
a fourth order ordinary linear differential equation, which has no singular properties 
if the Schwarzschild discriminant associated with the buoyant force does 
not vanish inside the star.
By solving a simplified version of the extended Kojima's equation for a simple
toy model,
Lockitch \& Andersson (2001) showed that because of the higher order rotational terms  
the singular solution in the original Kojima's equation can be avoided.
Obviously, we need to solve the complete version of the extended Kojima's equation to
obtain a definite conclusion concerning the continuous singular $r$-mode solutions.

It may be instructive to turn our attention to 
a difference in mathematical property between
the Newtonian $r$-modes 
and the relativistic $r$-modes associated with Kojima's equation. 
It is usually assumed that in the lowest order of $\Omega$
the eigenfunction as well as the eigenfrequency of the $r$-modes is proportional to $\Omega$.
In the case of the Newtonian $r$-modes, 
if we employ a perturbative method for $r$-modes in which the angular frequency
$\Omega$ is regarded as a small expanding parameter,
the radial eigenfunctions of order of $\Omega$ can be determined 
by solving a differential equation derived from the terms of order of $\Omega^3$,  
which bring about the couplings between the oscillations and the buoyant force
in the interior
(e.g., Provost et al 1981, Saio 1982).
In other words, there is no differential equation, 
of order of $\Omega$, which determines the
radial eigenfunction of the Newtonian $r$-modes.
On the other hand,
in the case of the general relativistic $r$-modes derived from Kojima's equation, 
we do not have to take account of the rotational effects of order of $\Omega^3$
to obtain the radial eigenfunctions of order of $\Omega$.
That is, the eigenfrequency and eigenfunction of the $r$-modes are both
determined by a differential equation (i.e., Kojima's equation) derived from 
the terms of order of $\Omega$.
This remains true even if we take the Newtonian limit of Kojima's equation
to calculate the $r$-modes (Lockitch et al. 2001; Yoshida 2001).
We think that this is an essential difference between the Newtonian $r$-modes
and the relativistic $r$-modes associated with Kojima's equation.
Considering that Kojima's equation can give no relativistic counterpart of the
Newtonian $r$-mode,
it is tempting to assume that some terms representing certain physical processes
are missing in the original Kojima' equation.
On the analogy of the $r$-modes in Newtonian dynamics, 
we think that the buoyant force in the interior plays an essential role to obtain
a relativistic counterpart of the Newtonian $r$-modes and 
that the terms due to the buoyant force
will appear when the rotational effects higher than the first order of $\Omega$
are included.
This is consistent with the suggestions made by
Kojima \& Hosonuma (2000) and Lockitch \& Andersson (2001).
In this paper, 
we calculate relativistic $r$-modes by taking account of the effects of the buoyant force
in a relativistic version of the Cowling approximation, in which all the 
metric perturbations are omitted.
In our formulation, all the terms associated with the Coriolis force are included but 
the terms from the centrifugal force are all ignored.
Our formulation can take account of the rotational contributions, due to the Coriolis force,
higher than the first order of $\Omega$.
Note that our method of solution is not a perturbation theory in which
$\Omega$ is regarded as a small expanding parameter for the eigenfrequency and eigenfunction.
Similar treatment has been employed in Newtonian stellar pulsations in rotating stars
(see, e.g., Lee \& Saio 1986; Unno et al. 1989;
Bildsten, Ushomirsky \& Cutler 1996; Yoshida \& Lee 2001). 
This treatment is justified
for low frequency modes because the Coriolis force 
dominates the centrifugal force in the equations of motion. 
The plan of this paper is as follows. In \S 2, we describe the formulation 
of relativistic stellar pulsations in the relativistic Cowling 
approximation. In \S 3, we show the modal properties of the $r$-modes in
neutron star models. In \S 4, we discuss about our results, specially the effect of 
the buoyancy in the fluid core on the modes. \S 5 is devoted to conclusions. 
In this paper, we use units in which $c=G=1$, where $c$ and $G$ denote the 
velocity of light and the gravitational constant, respectively.

%%%%%%%%%%%%%%%%%%%%%%%%%%%%%%%%%%%%%%%%%%%%%%%%%%%%%%%%%%%%%%
%%  Formulation                                             %%
%%%%%%%%%%%%%%%%%%%%%%%%%%%%%%%%%%%%%%%%%%%%%%%%%%%%%%%%%%%%%%

\section{Formulation}

\subsection{Equilibrium State}

We consider slowly and uniformly rotating relativistic stars in equilibrium.
If we take account of the rotational effects up to first order of $\Omega$,
the geometry in the stars can be described by the following line element 
(see, e.g. Thorne 1971): 
\begin{eqnarray}
ds^2 = g_{\alpha\beta}dx^\alpha dx^\beta = 
     - e^{2 \nu(r)} dt^2 + e^{2 \lambda(r)} dr^2 + r^2 d\theta^2 + 
           r^2 \sin^2 \theta d\varphi^2 - 
           2 \omega(r) \, r^2 \sin^2 \theta dt d\varphi  \, . 
\label{metric}
\end{eqnarray}
The fluid four-velocity in a rotating star is given by 
\begin{equation}
u^\alpha = \gamma(r,\theta) \, (t^\alpha + \Omega\,\varphi^\alpha) \, , 
\end{equation}
where $t^\alpha$ and $\varphi^\alpha$ stand for the timelike and rotational 
Killing vectors, respectively. 
Here the function $\gamma$ is chosen to satisfy 
the normalization condition $u^\alpha u_\alpha = -1$. 
If we consider the accuracy up to order of $\Omega$, the function $\gamma$ reduces to:
\begin{equation}
\gamma = e^{-\nu(r)} \, .  
\end{equation}
Once the
physical quantities of the star such as the pressure, $p(r)$, 
the mass-energy density, $\rho(r)$, and the metric function, $\nu(r)$, are given, 
the rotational effect on the metric, $\omega(r)$ can be obtained from a 
well-known numerical procedure (see, e.g., Thorne 1971).

\subsection{Pulsation Equations in the Cowling Approximation}

General relativistic pulsation equations are usually obtained by linearizing
Einstein's field equation.
The linearized Einstein equation contains perturbations associated with
the metric fluctuations and the fluid motions.
In this paper, to simplify the problem, we employ
a relativistic version of the Cowling approximation, 
in which all the metric perturbations are omitted in the 
pulsation equations (see McDermott, Van Horn, \& Scholl 1983, and Finn 1988).
The relativistic Cowling approximation is accurate enough 
for the $f$- and $p$-modes in non-rotating stars (Lindblom \& Splinter 1992).
It is also the case for the modes in slowly rotating stars (Yoshida \& Kojima 1997). 
The relativistic Cowling approximation is a good approximation 
for oscillation modes in which
the fluid motions are dominating over the metric fluctuations 
to determine the oscillation frequency.
Therefore, it is justified to employ the relativistic
Cowling approximation for the $r$-modes, for which the fluid motion is dominating.

If we employ the Cowling approximation, we can obtain our basic equations for pulsations
from the perturbed energy and momentum conservation laws:
\begin{eqnarray}
&&\delta\,(u^\alpha \nabla_\beta T^\beta_\alpha) = 0 \, , \ \ \ \ 
{\rm (energy\ conservation\ law)} 
\label{ena-eq} \\
&&\delta\,(q^\alpha_\gamma \nabla_\beta T^\beta_\alpha) = 0 \, , \ \ \ \
{\rm (momentum\ conservation\ law)} 
\label{mom-eq}
\end{eqnarray}
where $\nabla_\alpha$ is the covariant derivative associated with the metric, 
$T^\alpha_\beta$ is the energy-momentum tensor, and $q^\alpha_\beta$ is 
the projection tensor with respect to the fluid four-velocity. 
Here, $\delta Q$ denotes the Eulerian change in the physical quantity $Q$. 
In this paper, we adapt the adiabatic condition for the pulsation: 
\begin{eqnarray}
\Delta p = \frac{p\,\Gamma}{\rho+p}\,\Delta \rho \, ,
\label{ad-rel}
\end{eqnarray}
where $\Gamma$ is the adiabatic index defined as 
\begin{eqnarray}
\Gamma = \frac{\rho+p}{p}\,\left(\frac{\partial p}{\partial \rho}\right)_{ad}
 \, ,
\end{eqnarray}
and $\Delta Q$ stands for the Lagrangian change in the physical quantity 
$Q$. 
The relation between the Lagrangian and the Eulerian changes is given 
by the equation: 
\begin{eqnarray}
\Delta Q = \delta Q + \pounds_\zeta Q \, , 
\end{eqnarray}
where $\pounds_\zeta$ is the Lie derivative along the Lagrangian displacement vector
$\zeta^\alpha$, which is defined by the relation:
\begin{eqnarray}
\delta \hat{u}^{\alpha} = q^\alpha_\beta \delta u^\beta = 
q^\alpha_\beta (\pounds_u \zeta)^\beta \, .
\end{eqnarray}
Notice that we have 
$\delta \hat{u}^{\alpha} = \delta u^\alpha$ in the Cowling 
approximation. 
Because we are interested in pulsations of stationary rotating 
stars, we can assume that all the perturbed quantities have 
time and azimuthal dependence given by $e^{i\sigma t+i m \varphi}$, where $m$ is 
a constant integer and $\sigma$ is a constant frequency measured by an inertial 
observer at the spatial infinity. 
Because of this assumption, the relation 
between the Lagrangian displacement $\zeta^\alpha$ and the velocity 
perturbation $\delta \hat{u}^\alpha$ reduces to an algebraic relationship: 
\begin{eqnarray}
\delta \hat{u}^\alpha = i \gamma \bar{\sigma} \zeta^\alpha \, , 
\label{def-disp}
\end{eqnarray}
where $\bar{\sigma}$ is the frequency defined in the corotating frame defined as 
$\bar{\sigma}=\sigma + m \Omega$. 
Note that the gauge freedom in 
$\zeta^\alpha$ has been used to demand the relation 
$u_\alpha \zeta^\alpha =0$. 
By substituting equations (\ref{ad-rel}) and 
(\ref{def-disp}) into equations (\ref{ena-eq}) and (\ref{mom-eq}), we can 
obtain the perturbed energy equation,     
\begin{eqnarray}
\frac{1}{\gamma}\,\nabla_\alpha (\gamma \zeta^\alpha) + 
\frac{1}{p \Gamma}\,(\delta p + \zeta^\alpha \nabla_\alpha p) = 0 \, ,
\label{ena-eq2}
\end{eqnarray}
and the perturbed momentum equation, 
\begin{eqnarray}
- \gamma^2\bar{\sigma}^2 g_{\alpha\beta}\zeta^\beta +
2 i \gamma \bar{\sigma} \zeta^\beta \nabla_\beta u_\alpha + 
q_\alpha^\beta \nabla_\beta\,\left( \frac{\delta p}{\rho+p}\right) + 
\left( \frac{\delta p}{\rho+p}\,q_\alpha^\beta + \zeta^\beta\, 
\frac{\nabla_\alpha p}{\rho+p} \right) \, A_\beta = 0 \, ,
\label{mom-eq2} 
\end{eqnarray}
where $A_\alpha$ is the relativistic Schwarzschild discriminant defined by 
\begin{eqnarray}
A_\alpha = \frac{1}{\rho+p}\,\nabla_\alpha \rho - 
           \frac{1}{\Gamma p}\,\nabla_\alpha p \, .
\end{eqnarray}
Notice that equations (\ref{ena-eq2}) and (\ref{mom-eq2}) have been derived 
without the assumption of slow rotation. 
Physically 
acceptable solutions of equations (\ref{ena-eq2}) and (\ref{mom-eq2}) must 
satisfy boundary conditions at the center and the 
surface of the star. 
The surface boundary condition at $r=R$ is given by
\begin{eqnarray}
\Delta p = \delta p + \zeta^\alpha \nabla_\alpha p = 0 \, ,
\end{eqnarray}
and the inner boundary condition is 
that all the eigenfunctions are regular at the center ($r=0$).

On the analogy between general relativity and Newtonian gravity, 
the second term on the left hand side of equation (\ref{mom-eq2}) is 
interpreted as a relativistic counterpart of the Coriolis force. 
In our formulation, the terms due to the Coriolis 
like force are included in the perturbation equations, but the terms due to the 
centrifugal like force, which are proportional to $\Omega^2/(GM/R^3)$, are 
all ignored. 
In Newtonian theory of oscillations, this approximation is justified for low frequency modes 
satisfying the conditions $\left|2\Omega/\bar{\sigma}\right|\ge 1$ and 
$\Omega^2/(GM/R^3)\ll1$ (Lee \& Saio 1986; Unno et al. 1989; 
Bildsten, Ushomirsky \& Cutler 1996; Yoshida \& Lee 2001). 
In general relativity, we note that it is difficult to make a clear 
distinction between inertial forces such as the Coriolis force and  
the centrifugal force. 
From a physical point of view, however, it is also acceptable to use the approximation 
for low frequency oscillations in the background spacetime described by 
the metric (\ref{metric}).

The eigenfunctions are expanded in terms of spherical harmonic
functions $Y^m_l(\theta,\varphi)$ with different values of $l$ for a given 
$m$.  
The Lagrangian displacement, $\zeta^k$ and the pressure perturbation, 
$\delta p/(\rho+p)$ are expanded as
\begin{equation}
\zeta^r = r \sum_{l\geq\vert m \vert}^{\infty}  \, S_l(r)  
Y_l^m (\theta,\varphi) \, e^{i \sigma t} \,  , 
\label{xi-r}
\end{equation}
\begin{equation}
\zeta^\theta = \sum_{l,l'\geq\vert m \vert}^{\infty} 
\left\{ H_l (r) {\partial {Y_l^m (\theta,\varphi)}\over\partial\theta}  
- T_{l'} (r) \frac{1}{\sin \theta} \, 
{\partial{Y_{l'}^m (\theta,\varphi)}\over\partial\varphi} \right\} \, 
e^{i \sigma t} \, ,
\label{xi-th}
\end{equation}
\begin{equation}
\zeta^\varphi = {1\over\sin^2\theta}\,  
\sum_{l,l'\geq\vert m \vert}^{\infty} \left\{ H_l (r) 
{\partial {Y_l^m (\theta,\varphi)}\over\partial\varphi}
        + T_{l'} (r) \sin\theta\,{\partial {Y_{l'}^m (\theta,\varphi) }
            \over\partial\theta} \right\} \, e^{i \sigma t} \, , 
\label{xi-ph}
\end{equation}
\begin{equation}
\frac{\delta p}{\rho+p} = \sum_{l \geq\vert m \vert}^{\infty} \delta U_l (r) 
Y_l^m (\theta,\varphi) \, e^{i \sigma t} \, , 
\label{del-p}
\end{equation}
where $l=|m|+2k$ and $l'=l+1$ for even modes and $l=|m|+2k+1$ and $l'=l-1$
for odd modes where $k=0,~1,~2~\cdots$ (Regge \& Wheeler 1957; Thorne 1980).   
Here, even and odd modes are, respectively, characterized 
by their symmetry and antisymmetry of the eigenfunction with respect to the 
equatorial plane. 
Substituting the perturbed quantities 
(\ref{xi-r})--(\ref{del-p}) into linearized equations (\ref{ena-eq2}) and 
(\ref{mom-eq2}), we obtain an infinite system of coupled ordinary differential 
equations for the expanded coefficients. 
The details of our basic equations are given in the Appendix. 
Note that non-linear terms of
$q\equiv 2\bar\omega/\bar{\sigma}$ where $\bar\omega\equiv \Omega-\omega$
are kept in our basic equations.  
For numerical calculations, the infinite set of ordinary differential equations are 
truncated to be a finite set by discarding all the expanding coefficients 
associated with $l$ larger than $l_{\rm max}$, the value of which is determined 
so that the eigenfrequency and the eigenfunctions are well converged
as $l_{\rm max}$ increases (Yoshida \& Lee 2000a).

%%%%%%%%%%%%%%%%%%%%%%%%%%%%%%%%%%%%%%%%%%%%%%%%%%%%%%%%%%%%%%
%%  r-Modes of Neutron Star Models                          %%
%%%%%%%%%%%%%%%%%%%%%%%%%%%%%%%%%%%%%%%%%%%%%%%%%%%%%%%%%%%%%%

\section{$r$-Modes of Neutron Star Models}

The neutron star models that we use in this paper are the same as those
used in the modal analysis by McDermott, Van Horn, \& Hansen (1988).
The models are taken from the evolutionary sequences 
for cooling neutron stars calculated by Richardson 
et al. (1982), where the envelope structure is constructed 
by following Gudmundsson, Pethick \& Epstein (1983). 
These models are composed of a fluid core, a solid crust and a surface 
fluid ocean, and the interior temperature is finite and is not 
constant as a function of the radial distance $r$. 
The models are not barotropic and the Schwarzschild discriminant 
$\vert A\vert$ has finite values in the interior of the star. 
In order to avoid the complexity in the modal properties of relativistic $r$-modes
brought about by the existence of the solid crust in the models
(see Yoshida \& Lee 2001), we treat the whole
interior of the models as a fluid in the following modal analysis.

We computed frequency spectra of $r$-modes for the neutron star models called
NS05T7, NS05T8, and NS13T8 (see, McDermott et al. 1988). 
The physical properties such as
the total mass $M$, the radius $R$, the central 
density $\rho_c$, the central temperature $T_c$ and the relativistic factor 
$GM/c^2 R$ are summarized in Table 1 
(for other quantities, see McDermott et al. 1988).  
In Figures 1 and 2, scaled eigenfrequencies $\kappa\equiv\bar{\sigma}/\Omega$ 
of the $r$-modes of the three neutron star models are 
given as functions of 
$\hat\Omega\equiv\Omega/\sqrt{GM/R^3}$ for $m=2$ and $3$ cases, respectively. 
Here only the fundamental $r$-modes with $l'=m$ are considered because they 
are most important for the $r$-mode instability of neutron stars. 
We note that it is practically impossible to correctly calculate 
rotationally induced modes at $\hat{\Omega} \sim 0$ because of their coupling with 
high overtone $g$-modes having extremely low frequencies. 
From these figures, we can see that the scaled eigenfrequency $\kappa$ is 
almost constant as $\hat\Omega$ varies. 
In other words, the relation $\bar\sigma \sim \kappa_0 \Omega$ 
is a good approximation for the fundamental $r$-modes with $l'=m$, 
where $\kappa_0$ is a constant. 
Comparing the two frequency curves, which nearly overlap each other, 
for the models NS05T7 and NS05T8,
it is found that the detailed interior structure of the stars such 
as the temperature distribution $T(r)$ does not strongly affect
the frequency of the fundamental $r$-modes with $l'=m$. 
This modal property is the same as that found for the fundamental $l'=m$ $r$-modes 
in Newtonian dynamics (see, Yoshida \& Lee 2000b).
On the other hand, comparing the frequency curves for the models NS05T7 (NS05T8)
and NS13T8, we note that
the $r$-mode frequency of relativistic stars is strongly dependent
on the relativistic factor $GM/c^2R$ of the models. 
This is because the values of the effective rotation frequency 
$\bar\omega\equiv\Omega-\omega$ 
in the interior is strongly influenced by the relativistic factor. 
Similar behavior of the $GM/c^2R$ 
dependence of the $r$-mode frequency has been found in the analysis of Kojima's 
equation (Yoshida 2001; Ruoff \& Kokkotas 2001a).   
In Table 2, the values of $\kappa_0$ for the $r$-modes shown in Figures 1 and 2 are 
tabulated, where $\kappa_0$ are evaluated at $\hat\Omega=0.1$. 
The boundary values for the continuous part of the frequency spectrum 
for the $l'=m=2$ $r$-modes 
derived from Kojima's equation (Kojima 1998) are also listed in the same table. 
As shown by Kojima (1998) (see, also, Beyer \& Kokkotas 1999; Lockitch et al. 2001), 
if an $r$-mode falls in the frequency region bounded by the boundary values, 
Kojima's equation becomes singular and yields a continuous frequency spectrum and 
singular eigenfunctions as solutions. 
Although all the $r$-modes obtained in the present study are in the bounded 
frequency region, 
our numerical procedure shows that the $r$-modes obtained here are  
isolated and discrete eigenmodes.
In fact, no sign of continuous frequency spectrum appears in the present numerical
analysis (for a sign 
of the appearance of a continuous frequency spectrum in a numerical analysis, see
Schutz \& Verdaguer 1983).

In Figures 3 and 4, the eigenfunctions $i\,T_2$ 
for the $l'=m=2$ fundamental $r$-modes 
in the neutron star models NS05T8 and NS13T8 at $\hat{\Omega}=0.1$ are shown. 
We can confirm from these figures that the
eigenfunctions show no singular property, even though the  
frequencies are in the continuous part of the spectrum associated with Kojima's equation. 
These figures also show
that the fluid motion due to the $r$-modes is more confined 
near the stellar surface for NS13T8 than for NS05T8. 
This suggests that the eigenfunctions $i\,T_m$ tend to be confined to the stellar surface 
as the relativistic factor of the star increases. 
The same property appears in regular $r$-mode solutions derived 
from Kojima's equation (Yoshida 2001; Ruoff \& Kokkotas 2001a; 
Yoshida \& Futamase 2001). 
But, in the present case, the confinement of the eigenfunction for NS13T8 
is not so strong, even though the model NS13T8 is highly 
relativistic in the sense that the relativistic factor is as large as
$GM/c^2R=0.249$.

\section{Discussion}

In the neutron star 
models analyzed in the last section, the buoyant force is produced by 
thermal stratification in the star. 
However, as suggested by 
Reisenegger \& Goldreich (1992), the buoyant force in the core of neutron 
stars might become even stronger if the effect of 
the smooth change of the chemical composition 
of charged particles (protons and electrons) in the core is taken into account.
We thus think it legitimate to examine the effects of the enhanced
buoyant force on the modal properties of the relativistic $r$-modes.
Although our neutron star models do not provide 
enough information regarding the composition gradient, 
we may employ for this experiment an approximation formula for the 
Schwarzschild discriminant due to the composition gradient in the core
given by Reisenegger \& Goldreich (1992):
\begin{equation}
r A_r = 3.0\times 10^{-3}\,\left(\frac{\rho}{\rho_{nuc}}\right)
\frac{r}{\rho}\frac{d\rho}{dr} \, , 
\label{cma}
\end{equation} 
where $\rho_{nuc}$ denotes the nuclear density $\rho_{nuc}=2.8\times10^{14} 
\, {\rm g \, cm^{-3}}$. 
In Figure 5, Brunt-V\"ais\"al\"a 
frequency due to the thermal stratification (solid line) and that due to the composition
gradient (dashed line) are given as a function of $\log(1-r/R)$ 
for model NS13T8, where the frequencies are normalized by $(GM/R^3)^{1/2}$.
In this paper, the relativistic Brunt-V\"ais\"al\"a frequency $N$ is defined as  
\begin{equation}
N^2 = \frac{A_r}{\rho+p}\, \frac{d p}{dr} \, .
\end{equation}
This figure shows that in the core
the Brunt-V\"ais\"al\"a frequency due to the composition gradient
is by several orders of magnitude larger and hence
the corresponding buoyant force is much stronger than those produced by
the thermal stratification.
Note that the Brunt-V\"ais\"al\"a frequency due to the composition gradient
becomes as large as $\sim(GM/R^3)^{1/2}$ in the core (see also, e.g., Lee 1995).

Replacing the Schwarzschild discriminant in the original neutron star models
with that due to the composition gradient calculated by using equation (19),
we computed the fundamental $r$-modes with $l'=m=2$ for the models NS05T8 and NS13T8,
and plotted $\kappa=\bar\sigma/\hat \Omega$
against rotation frequency $\hat{\Omega}$ in Figure 6, where
$\kappa$'s for the original models with the thermal stratification were
also plotted for convenience. 
As shown by the figure, $\kappa$ slightly increases
as $\hat\Omega\rightarrow 0$ for the models with the enhanced buoyant force.
This is remarkable because such behavior of $\kappa$ for the fundamental $r$-modes
with $l'=m$
is not found in the case of Newtonian 
$r$-modes in non-barotropic stars (Yoshida \& Lee 2000b).
The slight increase of $\kappa$ with decreasing $\hat\Omega$
may be explained in terms of the
property of the eigenfunctions of the $r$-modes.
We depict
the eigenfunctions $i\,T_2$ of the $l'=m=2$ fundamental $r$-modes 
in the neutron star model NS13T8 at $\hat{\Omega}=0.2$ and $.02$
in Figures 7 and 8,
where the solid line and dashed line in each figure denote the $r$-modes 
in the models with the compositional stratification
and with the thermal stratification, respectively.
As shown by Figure 7, in the case of rapid rotation,
the buoyant force in the core does not strongly
affect the properties of the eigenfunctions.
On the other hand, in the case of slow rotation, Figure 8 shows that
the amplitude of the eigenfunction is strongly confined to the region near
the surface for the model with the enhanced buoyant force.
For relativistic $r$-modes, as a result of the general relativistic frame dragging effect,
the effective rotation frequency $\bar\omega=\Omega-\omega\propto\Omega$ 
acting on local fluid elements 
is an increasing function of the distance $r$ from the stellar center, and
thus fluid elements near the stellar surface are rotating faster 
than those near the stellar center. 
The oscillation frequency of the $r$-modes may be determined by
the mean rotation frequency obtained by averaging local 
rotation frequencies over the whole interior of the star with
a certain weighting function associated with the eigenfunction.
In this case, it is reasonable to expect that 
the relativistic $r$-modes that have large amplitudes only in the regions
near the surface
get larger values of $\kappa$ than those that have large amplitudes deep in the core.

For the models with the enhanced buoyant force,
the current quadrupole moment of the $r$
modes that determines the strength of the instability will be largely reduced
when $\hat\Omega$ is small as a result of the strong confinement of the amplitudes.
But, since the $r$-mode instability is believed to operate at rapid rotation rates,
the effect of the amplitude confinement at small rotation rates
may be irrelevant to the instability.

%%%%%%%%%%%%%%%%%%%%%%%%%%%%%%%%%%%%%%%%%%%%%%%%%%%%%%%%%%%%%%
%%       DISCUSSIONS AND CONCLUSIONS                        %%
%%%%%%%%%%%%%%%%%%%%%%%%%%%%%%%%%%%%%%%%%%%%%%%%%%%%%%%%%%%%%%

\section{Conclusion}

In this paper, we have investigated the properties of relativistic
$r$-modes in slowly rotating neutron stars 
in the relativistic Cowling approximation by taking account of 
higher order effects of rotation than the first oder of $\Omega$.
In our formalism, only the influence of the Coriolis like 
force on the oscillations are taken into account, and no effects of 
the centrifugal like force are considered. 
We obtain the fundamental $r$-modes associated with 
$l'=m=2$ and $3$ for three neutron star models. 
We find that the fundamental $r$-mode frequencies are in a good approximation 
given by $\bar\sigma\approx \kappa_0\,\Omega$.
The proportional coefficient $\kappa_0$ 
is only weakly dependent on $\Omega$, but strongly
depends on the relativistic parameter $GM/c^2R$.
For the fundamental $r$-modes with $l'=m$, we find that
the buoyant force in the core is more influential to the relativistic
$r$-modes than to the Newtonian $r$-modes.
All the $r$-modes obtained in this paper are discrete modes with distinct regular 
eigenfunctions, and they all fall in 
the frequency range of the continuous spectrum of Kojima's equation. 
We may conclude that the fundamental $r$-modes obtained in this paper are
the relativistic counterpart of the Newtonian $r$-modes.

Here, it is legitimate to mention the relation between the present work and 
the study by Kojima \& Hosonuma (1999), who studied relativistic $r$-modes 
in slowly rotating stars in the Cowling approximation. 
By applying the Laplace transformation to linearized equations 
derived for the $r$-modes, Kojima \& Hosonuma (1999) examined 
how a single component of initial perturbations with axial parity
evolves as time goes, and showed
that the perturbations cannot oscillate with a single frequency. 
In this paper, we have presented a formulation for small
amplitude relativistic oscillations in rotating stars 
in the relativistic Cowling approximation,
and solved the oscillation equations as a boundary-eigenvalue problem for the $r$-modes.
In our formulation, we assume neither axially dominant eigenfunctions nor
low frequencies to calculate the $r$-modes.
Considering these differences in the treatment of the $r$-mode oscillations
between the two studies,
it is not surprising that our results are not necessarily consistent with
those by Kojima \& Hosonuma (1999).

Our results suggest that the appearance of singular $r$-mode solutions
can be avoided by extending the original Kojima's equation so that
terms due to the buoyant force in the stellar interior are included. 
As discussed recently by Yoshida \& Futamase (2001) and Lockitch \& 
Andersson (2001), we believe that the answer to the question whether 
$r$-mode oscillations in uniformly rotating relativistic stars show true 
singular behavior may be given by solving the forth order 
ordinary differential equation derived by Kojima \& 
Hosonuma (2000) for $r$-modes. 
Verification of this possibility remains as a future study.

%%%%%%%%%%%%%%%%%%%%%%%%%%%%%%%%%%%%%%%%%%%%%%%%%%%%%%%%%%%%%%
%%         END OF MAIN BODY OF PAPER                        %%
%%%%%%%%%%%%%%%%%%%%%%%%%%%%%%%%%%%%%%%%%%%%%%%%%%%%%%%%%%%%%%

\acknowledgements

We would like to thank H. Saio for useful comments. We are grateful to the 
anonymous referee for useful suggestions regarding the buoyancy due to the 
composition gradient. S.Y. would like to thank Y. Eriguchi and T. Futamase for 
fruitful discussions and continuous encouragement.

%%%%%%%%%%%%%%%%%%%%%%%%%%%%%%%%%%%%%%%%%%%%%%%%%%%%%%%%%%%%%%
%% APPENDIX                                                 %%
%%%%%%%%%%%%%%%%%%%%%%%%%%%%%%%%%%%%%%%%%%%%%%%%%%%%%%%%%%%%%%

\appendix

%%%%%%%%%%%%%%%%%%%%%%%%%%%%%%%%%%%%%%%%%%%%%%%%%%%%%%%%%%%%%%
%% Basic equations for ...                                  %%
%%%%%%%%%%%%%%%%%%%%%%%%%%%%%%%%%%%%%%%%%%%%%%%%%%%%%%%%%%%%%%

\section{Basic Equations}

We introduce column vectors $\bf{y}_1$, $\bf{y}_2$, $\bf{h}$, and $\bf{t}$, 
whose components are defined by 
\begin{equation}
y_{1, k} = S_l (r) \, ,
\end{equation}
\begin{equation}
y_{2, k} = \frac{1}{r \frac{d\nu}{dr}} \,  \delta U_l(r) \, , 
\end{equation}
\begin{equation}
h_{,k} = H_l (r) \, ,
\end{equation}
and
\begin{equation}
t_{,k} = T_{l'} (r) \, ,
\end{equation}
where $l=\vert m \vert + 2 k -2 $ and $l'=l+1$ for ``even'' modes, and 
$l=\vert m \vert + 2 k -1 $ and $l'=l-1$ for ``odd'' modes, and $k = 1, 2, 3, \dots$.

In vector notation, equations for the adiabatic nonradial pulsation
in a slowly rotating star are written as follows:

\noindent
The perturbed energy equation (\ref{ena-eq2}) reduces to 
\begin{eqnarray}
r \, \frac{d {\bf y}_1}{dr} + 
\left( 3 - \frac{V}{\Gamma} + r\frac{d \lambda}{dr} \right) \, {\bf y}_1
+ \frac{V}{\Gamma}\,{\bf y}_2 - {\bf \Lambda}_0 {\bf h}  
+c_2 \hat{\bar{\omega}} \hat{\sigma} \, (-m {\bf h}+{\bf C}_0 i {\bf t}) 
= 0 \, .
\label{differ-1}
\end{eqnarray}
The $r$ component of the perturbed momentum equations (\ref{mom-eq2}) 
reduces to 
\begin{eqnarray}
r \, \frac{d{\bf y}_2}{dr} - 
( e^{2\lambda}\,c_1 \hat{\bar{\sigma}}^2 + r A_r ) \, {\bf y}_1 
+(U+r A_r) {\bf y}_2 - c_1 \hat{\bar{\sigma}}^2 \chi
(-m {\bf h}+{\bf C}_0 i {\bf t}) = 0 \, .
\label{differ-2}
\end{eqnarray}
Here, 
\begin{equation}
U = \frac{r\frac{d}{dr}(r\frac{d\nu}{dr})}{\frac{d\nu}{dr}} \, , 
\hspace{0.5in}
V = - \frac{d\ln p}{d\ln r} \, ,
\end{equation}
\begin{equation}
\chi = \frac{e^{2\nu}}{r}\frac{d}{dr} 
\left( r^2 e^{-2 \nu} \frac{\bar{\omega}}{\bar{\sigma}}\right)\, ,
\end{equation}
\begin{equation}
c_1 = \frac{r^2 e^{-2 \nu}}{r \frac{d\nu}{dr}}\, \frac{M}{R^3} \, , 
\hspace{0.5in}
c_2 = r^2 e^{-2 \nu} \frac{M}{R^3} \, ,
\end{equation}
and
$\hat{\bar{\omega}} \equiv \bar{\omega}/(GM/R^3)^{1/2}$, 
$\hat{\bar{\sigma}} \equiv \bar{\sigma}/(GM/R^3)^{1/2}$, and 
$\hat{\sigma} \equiv \sigma/(GM/R^3)^{1/2}$ are frequencies in the unit of 
the Kepler frequency at the stellar surface, where $\bar\omega\equiv\Omega-\omega$
and $\bar\sigma\equiv\sigma+m\Omega$. 

\noindent
The $\theta$ and $\varphi$ components of the perturbed momentum equations 
(\ref{mom-eq2}) reduce to 
\begin{equation}
{\bf L}_0 \, {\bf h} + {\bf M}_1 \, i {\bf t}
 = \frac{1}{c_1 \hat{\bar{\sigma}}^2} \, {\bf y}_2 + \lbrace 
{\bf O}+{\bf M}_1{\bf L}_1^{-1}{\bf K}
 \rbrace \, \left(\chi \,{\bf y}_1+
 \frac{c_2 \hat{\bar{\omega}}\hat{\sigma}}{c_1 \hat{\bar{\sigma}}^2}\,
 {\bf y_2} \right) \, , \label{alge-1}
\end{equation}
\begin{equation}
{\bf L}_1 \, i {\bf t} + {\bf M}_0 \, {\bf h}
 = {\bf K} \, \left(\chi \,{\bf y}_1+
 \frac{c_2 \hat{\bar{\omega}}\hat{\sigma}}{c_1 \hat{\bar{\sigma}}^2}\,
 {\bf y_2} \right) \, , \label{alge-2}
\end{equation}
where 
\begin{equation}
{\bf O} = m {\bf \Lambda}_0^{-1} - {\bf M}_1{\bf L}_1^{-1}{\bf K} \, . 
\end{equation}

The quantities ${\bf C}_0$, ${\bf K}$, ${\bf L}_0$, 
${\bf L}_1$, ${\bf \Lambda}_0$, ${\bf \Lambda}_1$, ${\bf M}_0$, 
${\bf M}_1$ are matrices written as follows:  

\noindent
For even modes, 
\[
({\bf C}_0)_{i,i} = - (l+2) J^m_{l+1} \, , \hspace{.3in}
({\bf C}_0)_{i+1,i} = (l+1) J^m_{l+2} \, ,
\]
\[
({\bf K})_{i,i} = \frac{J^m_{l+1}}{l+1} \, , \hspace{.3in}
({\bf K})_{i,i+1} = - \frac{J^m_{l+2}}{l+2} \, ,
\]
\[
({\bf L}_0)_{i,i} = 1 - \frac{m q}{l(l+1)} \, , \hspace{.3in}
({\bf L}_1)_{i,i} = 1 - \frac{m q}{(l+1)(l+2)} \, ,
\]
\[
({\bf \Lambda}_0)_{i,i} = l(l+1) \, , \hspace{.3in}
({\bf \Lambda}_1)_{i,i} = (l+1)(l+2) \, ,
\]
\[
({\bf M}_0)_{i,i} = q \frac{l}{l+1} \, J^m_{l+1} \, , \hspace{.3in}
({\bf M}_0)_{i,i+1} = q \frac{l+3}{l+2} \, J^m_{l+2} \, ,
\]
\[
({\bf M}_1)_{i,i} = q \frac{l+2}{l+1} \, J^m_{l+1} \, , \hspace{.3in}
({\bf M}_1)_{i+1,i} = q \frac{l+1}{l+2} \, J^m_{l+2} \, ,
\]
where $l=\vert m \vert + 2 i -2 $ for $i = 1,2,3,\dots $, and
$q \equiv 2 \bar{\omega} / \bar{\sigma}$, and
\begin{equation}
J^m_l \equiv \left[ \frac{(l+m)(l-m)}{(2l-1)(2l+1)} \right]^{1/2} 
\, .  \label{J_l}
\end{equation}

\noindent
For odd modes, 
\[
({\bf C}_0)_{i,i} = (l-1) J^m_{l} \, , \hspace{.3in}
({\bf C}_0)_{i,i+1} = -(l+2) J^m_{l+1} \, ,
\]
\[
({\bf K})_{i,i} = - \frac{J^m_l}{l} \, , \hspace{.3in}
({\bf K})_{i+1,i} = \frac{J^m_{l+1}}{l+1} \, ,
\]
\[
({\bf L}_0)_{i,i} = 1 - \frac{m q}{l(l+1)} \, , \hspace{.3in}
({\bf L}_1)_{i,i} = 1 - \frac{m q}{l(l-1)} \, ,
\]
\[
({\bf \Lambda}_0)_{i,i} = l(l+1) \, , \hspace{.3in}
({\bf \Lambda}_1)_{i,i} = l(l-1) \, ,
\]
\[
({\bf M}_0)_{i,i} = q \frac{l+1}{l} \, J^m_l \, , \hspace{.3in}
({\bf M}_0)_{i+1,i} = q \frac{l}{l+1} \, J^m_{l+1} \, ,
\]
\[
({\bf M}_1)_{i,i} = q \frac{l-1}{l} \, J^m_l \, , \hspace{.3in}
({\bf M}_1)_{i,i+1} = q \frac{l+2}{l+1} \, J^m_{l+1} \, ,
\]
where $l=\vert m \vert + 2 i -1 $ for $i = 1,2,3,\dots $.

Eliminating $\bf h$ and 
$\it i \bf t$ from equations (\ref{differ-1}) and (\ref{differ-2}) by 
using equations (\ref{alge-1}) and (\ref{alge-2}), equations 
(\ref{differ-1}) and (\ref{differ-2}) reduce to a set of first-order 
linear ordinary  differential equations for ${\bf y}_1$ and ${\bf y}_2$ 
as follows:
\begin{eqnarray}
r\,\frac{d {\bf y}_1}{dr} &=& 
 \left\lbrace \left( \frac{V}{\Gamma} -3 -r\,\frac{d\lambda}{dr}\right) \, 
 {\bf 1} + \chi \,\left({\bf{\cal F}}_{11} + 
 c_2\hat{\bar{\omega}}\hat{\sigma} 
 {\bf{\cal F}}_{21}\right) \right\rbrace \, {\bf y}_1 \nonumber \\
&+& \left\lbrace 
\frac{1}{c_1 \hat{\bar{\sigma}}^2}\left({\bf{\cal F}}_{12} + 
 c_2\hat{\bar{\omega}}\hat{\sigma}
 {\bf{\cal F}}_{22}\right) + 
 \frac{c_2 \hat{\bar{\omega}}\hat{\sigma}}{c_1 \hat{\bar{\sigma}}^2}
 \left({\bf{\cal F}}_{11} + c_2\hat{\bar{\omega}}\hat{\sigma}
 {\bf{\cal F}}_{21}\right) - \frac{V}{\Gamma}{\bf 1} 
\right\rbrace 
\, {\bf y}_2 \, , \label{basic-eq1}
\end{eqnarray}
\begin{eqnarray}
r\,\frac{d {\bf y}_2}{dr} &=& \Bigl\lbrace 
(e^{2\lambda} c_1 \hat{\bar{\sigma}}^2 + r A_r) \, {\bf 1} - 
c_1 \hat{\bar{\sigma}}^2 q^2 {\bf{\cal F}}_{21} \Bigr\rbrace \, 
{\bf y}_1 \nonumber \\ 
&+& \Bigl\lbrace -(r A_r+U){\bf 1} - \chi \,( {\bf{\cal F}}_{22} + 
 c_2\hat{\bar{\omega}}\hat{\sigma} {\bf{\cal F}}_{21} )\Bigr\rbrace 
 \, {\bf y}_2 \, , \label{basic-eq2}
\end{eqnarray}
where 
\begin{equation}
{\bf{\cal F}}_{11} = {\bf W} {\bf O} \, ,
\end{equation}
\begin{equation}
{\bf{\cal F}}_{12} = {\bf W} \, ,
\end{equation}
\begin{eqnarray}
{\bf{\cal F}}_{21} = {\bf{\cal R}} {\bf W} {\bf O} 
 - {\bf C}_0 {\bf L}_1^{-1} {\bf K} \, ,
\end{eqnarray}
\begin{equation}
{\bf{\cal F}}_{22} = {\bf{\cal R}} \, {\bf W} \, ,
\end{equation}
\begin{eqnarray}
{\bf{\cal R}} = m{\bf \Lambda}_0^{-1} 
 + {\bf C}_0 {\bf L}_1^{-1} {\bf M}_0 {\bf \Lambda}_0^{-1} \, ,
\end{eqnarray}
\begin{equation}
{\bf W} = {\bf \Lambda}_0 
({\bf L}_0 - {\bf M}_1 {\bf L}_1^{-1} {\bf M}_0 )^{-1} \, .
\end{equation}
The surface boundary conditions $\Delta p (r=R)  = 0$ are given by 
\begin{equation}
{\bf y}_1 - {\bf y}_2  = 0 \, . \label{boundary-1}
\end{equation}
The inner boundary conditions at the stellar center are the regularity 
conditions of the eigenfunctions.

%%%%%%%%%%%%%%%%%%%%%%%%%%%%%%%%%%%%%%%%%%%%%%%%%%%%%%%%%%%%%%
%%  BIBLIOGRAPHY                                            %%
%%%%%%%%%%%%%%%%%%%%%%%%%%%%%%%%%%%%%%%%%%%%%%%%%%%%%%%%%%%%%%

%

%%%%%%%%%%%%%%%%%%%%%%%%%%%%%%%%%%%%%%%%%%%%%%%%%%%%%%%%%%%%%%
%% TABLES                                                   %%
%%%%%%%%%%%%%%%%%%%%%%%%%%%%%%%%%%%%%%%%%%%%%%%%%%%%%%%%%%%%%%

\newpage

\begin{deluxetable}{cccccc}
\footnotesize
\tablecaption{Neutron Star Models}
\tablewidth{0pt}
\tablehead{ \colhead{Model} & \colhead{$M\ (M_{\sun})$}  
 & \colhead{$R$ (km)} & \colhead{$\rho_c$ (g $\rm cm^3$)}    
 & \colhead{$T_c$ (K)} & \colhead{$GM/(c^2 R)$}
} 
\startdata
NS05T7&$0.503$&$9.839$&$9.44\times 10^{14}$&$1.03\times 10^7$
&$7.54\times 10^{-2}$\nl
NS05T8&$0.503$&$9.785$&$9.44\times 10^{14}$&$9.76\times 10^7$
&$7.59\times 10^{-2}$\nl 
NS13T8&$1.326$&$7.853$&$3.63\times 10^{15}$&$1.05\times 10^8$
&$2.49\times 10^{-1}$\nl
\enddata
\label{ns-model}
\end{deluxetable}

\begin{deluxetable}{ccccc}
\footnotesize
\tablecaption{Scaled Eigenfrequencies $\kappa_0$ of Fundamental 
$r$-modes}
\tablewidth{0pt}
\tablehead{
 \colhead{Model}&\colhead{$\kappa_0 (l=m=2)$}
                &\colhead{$\kappa_0 (l=m=3)$}
                &\colhead{$2/3\times\bar{\omega}(0)/\Omega$}
                &\colhead{$2/3\times\bar{\omega}(R)/\Omega$}
} 
\startdata
NS05T7 &$ 0.600 $&$ 0.455 $&$ 0.523 $&$ 0.645 $ \nl
NS05T8 &$ 0.601 $&$ 0.456 $&$ 0.524 $&$ 0.642 $ \nl
NS13T8 &$ 0.393 $&$ 0.309 $&$ 0.208 $&$ 0.489 $ \nl
\enddata
\label{eigen-f}
\end{deluxetable}

%%%%%%%%%%%%%%%%%%%%%%%%%%%%%%%%%%%%%%%%%%%%%%%%%%%%%%%%%%%%%%
%% FIGURES                                                  %%
%%%%%%%%%%%%%%%%%%%%%%%%%%%%%%%%%%%%%%%%%%%%%%%%%%%%%%%%%%%%%%

\newpage

\begin{figure}
\epsscale{.5}
\plotone{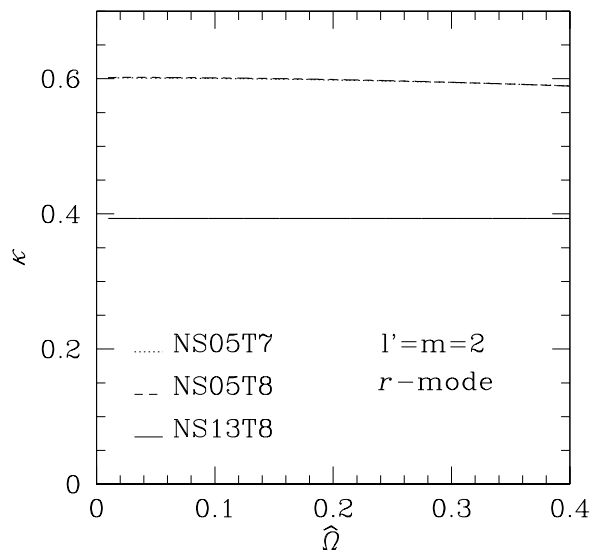}
\caption{Scaled frequencies $\kappa=\bar{\sigma}/\Omega$ of 
the fundamental $r$-modes in the neutron star models NS05T7, NS05T8, and NS13T8 
are plotted as functions of $\hat{\Omega}=\Omega/(GM/R^3)^{1/2}$ for 
$l'=m=2$. 
Note that the two $r$-mode frequency curves for the models NS05T7 and NS05T8
overlap each other almost completely.}
\end{figure}

%\newpage

\begin{figure}
\epsscale{.5}
\plotone{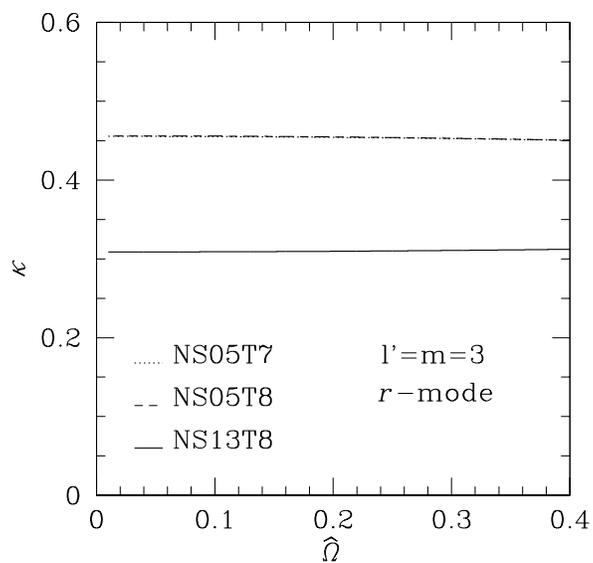}
\caption{Same as Figure 1 but for the case of $m=3$.}
\end{figure}

\newpage

\begin{figure}
\epsscale{.5}
\plotone{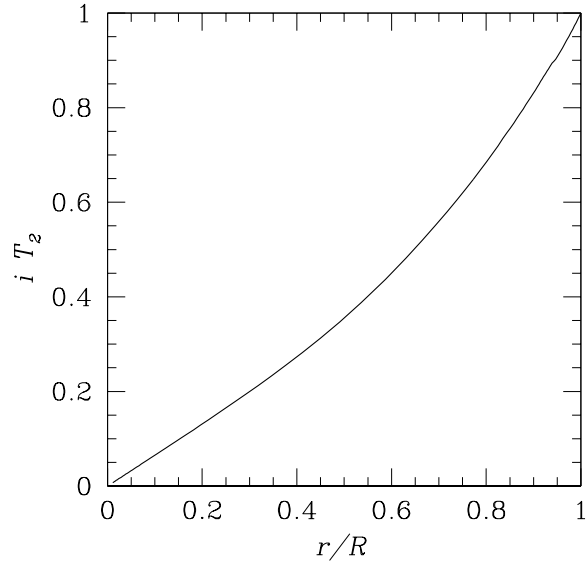}
\caption{Eigenfunction $i\ T_2$ of the 
$r$-mode with $l'=m=2$ for the model NS05T8 at $\hat{\Omega}=0.1$ 
is given as a function of $r/R$. Here, normalization of the eigenfunction 
is chosen as $i\ T_2(R)=1$.} 
\end{figure}

\begin{figure}
\epsscale{.5}
\plotone{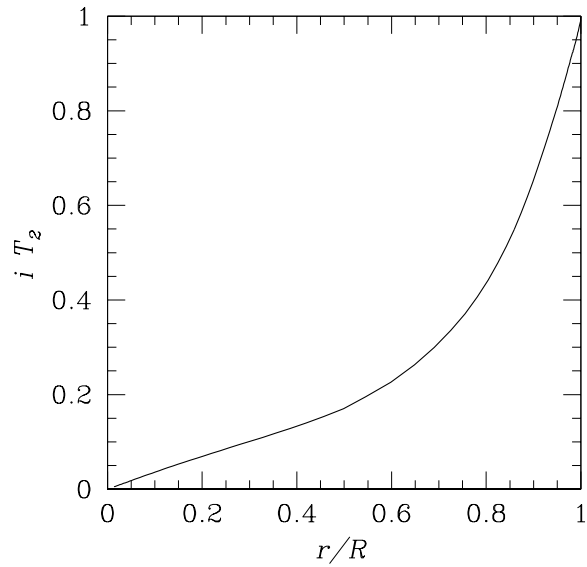}
\caption{Same as Figure 4 but for the model NS13T8.}
\end{figure}

\begin{figure}
\epsscale{.5}
\plotone{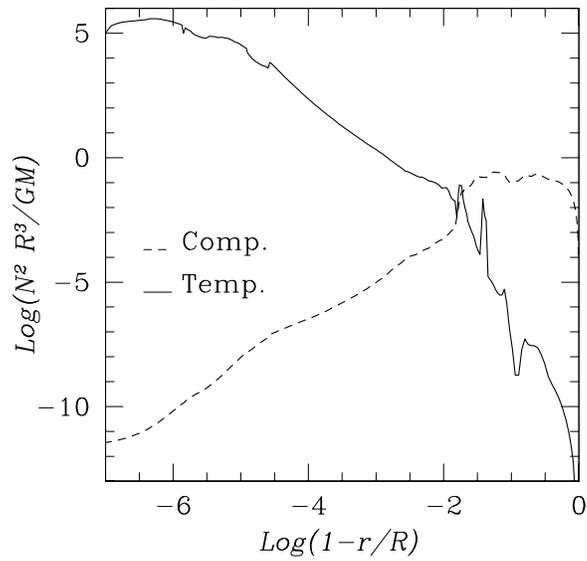}
\caption{Brunt-V\"ais\"al\"a frequency due to the thermal stratification (solid line) and that 
due to the composition gradient (dashed line) are given as a function of $\log(1-r/R)$ 
for model NS13T8, where the frequencies are normalized by $(GM/R^3)^{1/2}$.}
\end{figure}

\begin{figure}
\epsscale{.5}
\plotone{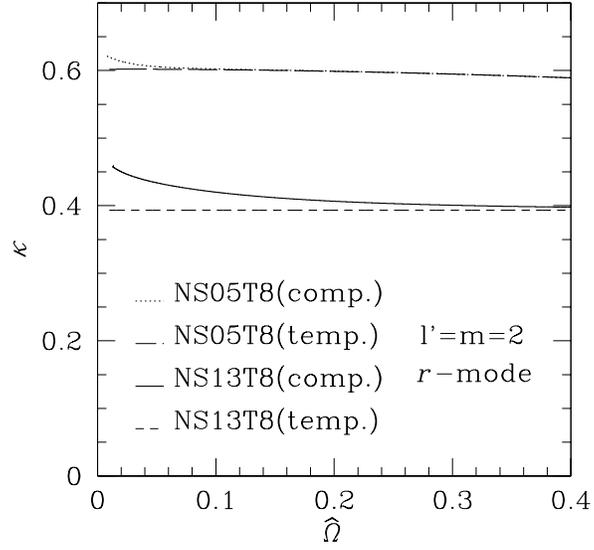}
\caption{Scaled frequencies $\kappa=\bar{\sigma}/\Omega$ of 
the $l'=m=2$ fundamental $r$-modes in the neutron star models NS05T8 and NS13T8 
are plotted as functions of $\hat{\Omega}=\Omega/(GM/R^3)^{1/2}$. 
Labels "comp." and "temp." stand for the frequencies of the models with 
the buoyancy due to the composition gradient and due to the temperature gradient, 
respectively.}
\end{figure}

\begin{figure}
\epsscale{.5}
\plotone{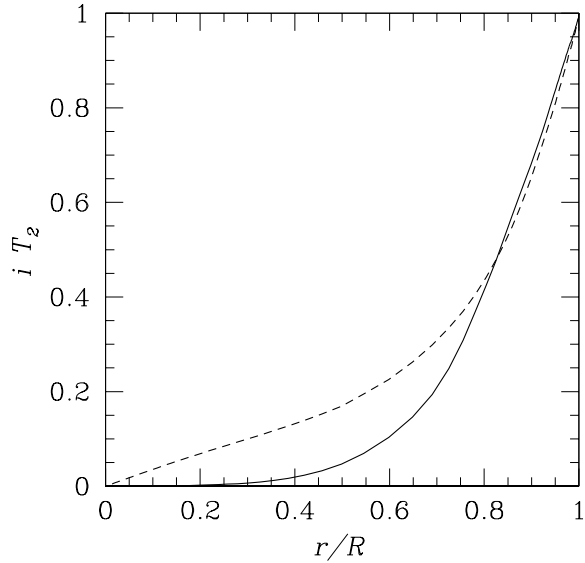}
\caption{Eigenfunctions $i\,T_2$ of the $l'=m=2$ fundamental $r$-modes 
in the neutron star model NS13T8 at $\hat{\Omega}=0.2$ are given as a function 
of $r/R$. Here, normalization of the eigenfunction is chosen as $i\ T_2(R)=1$.
The solid line and dashed line denote the $r$-modes in the models with the compositional 
stratification and with the thermal stratification, respectively.}  
\end{figure}

\begin{figure}
\epsscale{.5}
\plotone{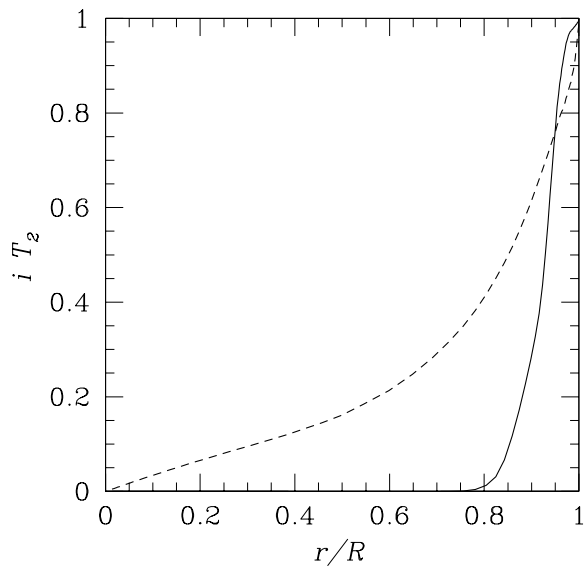}
\caption{Same as Figure 7 but for the model at $\hat{\Omega}=0.02$.}
\end{figure}

\end{document}